\documentclass[twocolumn,showpacs,floats,floatfix,aps,pra]{revtex4}
\usepackage{amsfonts}
\usepackage{amssymb}
\usepackage{amsmath}
\usepackage{calc,graphicx,color,bm}

\def\be{\begin{equation}}
\def\ee{\end{equation}}
\def\bea{\begin{eqnarray}}
\def\eea{\end{eqnarray}}
\def\bse{\begin{subequations}}
\def\ese{\end{subequations}}

\def\e{\,\text{e}}

\def\to{\rightarrow}
\def\fromto{\leftrightarrow}
\def\opt{o}

\def\ket#1{\vert #1\rangle}
\def\bra#1{\langle #1\vert}

\def\U{\mathbf{U}}

\def\phase{\phi}
\def\compphase{\varphi}

\def\phonons{\mathrm{v}}
\def\phonons{v}
\def\Toff{\mathbf{U}_\mathrm{T}}
\def\Toff{\mathbf{C}_n}
\def\ToffT{\widetilde{U}_{\mathrm{T}}^\dagger}
\def\ToffT{\mathbf{S}_n}
\def\nc{m}

\def\etal{\textit{et al.}}

\begin{document}

\author{Svetoslav S. Ivanov}
\affiliation{Department of Physics, Sofia University, 5 James Bourchier blvd, 1164 Sofia, Bulgaria}
\author{Nikolay V. Vitanov}
\affiliation{Department of Physics, Sofia University, 5 James Bourchier blvd, 1164 Sofia, Bulgaria}
\title{Scalable uniform construction of highly-conditional quantum gates}
\date{\today }

\begin{abstract}
We present a scalable uniform technique for construction of highly conditional C$^n$-NOT quantum gates of trapped ion qubits, such as the Toffoli gate, without using ancilla states and circuits of an exorbitant number of concatenated one- and two-qubit gates.
Apart from the initial dressing of the internal qubit states with vibrational phonons and the final restoration of the phonon ground state,
our technique requires the application of just a single composite pulse on the target qubit and is applicable both in and outside the Lamb-Dicke regime. We design special narrowband composite pulses, which suppress all transitions but the conditional transition of the target qubit;
moreover, these composite pulses significantly improve the spatial addressing selectivity.
\end{abstract}


\maketitle

\section{Introduction}\label{introduction}
%%%%%%%%%%%%%%%%%%%%%%%%%%%%%%%%%%%%%%%%%%%%%%%%%%%%%%%%%%%%%%%%%%%%%%%%%%%%%%%%%%%%%%%%%%%%%%%%%%%%%%%%%%%%%%%%%%%%%%%
One of the most important highly-conditional quantum gates is the three qubit control-control-NOT %(C$^2$-NOT)
 gate, known as the Toffoli gate,
 in which the target qubit is inverted if both control qubits are in state $\ket{1}$, and is left unchanged otherwise.
This gate has a central role in quantum error correction \cite{Cory1998}; %, which is vital for reliable quantum information processing.
 moreover, it forms with the one-qubit Hadamard gate a universal set of quantum gates \cite{Shi}.
The more general C$^n$-NOT gates are often used in quantum computing, e.g. as oracles in Grover's search \cite{Chuang2000} and to simulate quantum walks \cite{Hines2007}.

The simplest decomposition of the Toffoli gate in the circuit model of quantum computation uses six CNOT gates \cite{Chuang2000},
 or five CNOT gates with an ancilla state \cite{NPhys2009}.
Extending this approach to C$^n$-NOT gates with $n>2$ is highly demanding for it requires the ability to construct, with very high fidelity, many concatenated gates of this type.
It is hence desirable to seek simpler schemes for conditional multi-qubit gates without sequences of CNOT gates \cite{Borrelli}.
Recently, Monz \etal\ \cite{Blatt2009} have demonstrated experimentally the Toffoli gate with 71\% fidelity with ${}^{40}$Ca$^+$ ions by using a sophisticated sequence of 15 laser pulses in the Lamb-Dicke (LD) regime;
 it is not obvious, however, how this approach can be extended to higher C$^n$-NOT gates and outside the LD regime.

In this paper, we propose a simplified uniform scheme for construction of C$^n$-NOT gates of arbitrary order $n$ in a linear ion string by using specially designed composite pulses. The method does not use ancilla states and circuits of concatenated one- and two-qubit gates, and is applicable both inside and outside the LD regime. We design special narrowband (NB) composite pulses whose excitation profiles allow us to drive only transitions between a pair of selected collective ionic states,
 thereby manipulating the target qubit in a way, controlled by the other qubits.
The method allows us to construct C$^n$-NOT gates of various orders $n$ with essentially the same composite sequences, and hence the same level of complexity.

%%%%%%%%%%%%%%%%%%%%%%%%%%%%%%%%%%%%%%%%%%%%%%%%%%%%%%%%%%%%%%%%%%%%%%%%%%%%%%%%%%%%%%%%%%%%%%%%%%%%%%%%%%%%%%%%%%%%%%%
\section{Basic steps}\label{steps}
Our method begins with the initialization of the string of $n+1$ ions in the collective vibrational ground state $\ket{\phonons=0}$ \cite{comment1}.
We adopt the wavefunction notation $\ket{\psi} \ket{\phonons}$, where $\ket{\psi}=\ket{q_1q_2\cdots q_{n+1}}$ is the collective internal state of the ion qubits, with $q_k=0$ or 1,
 and $\ket{\phonons}$ is the collective phonon state.
The C$^n$-NOT gate is constructed in 5 steps.

%====================================================================================
\begin{figure}[t]\centering
    \includegraphics[width=0.90\columnwidth]{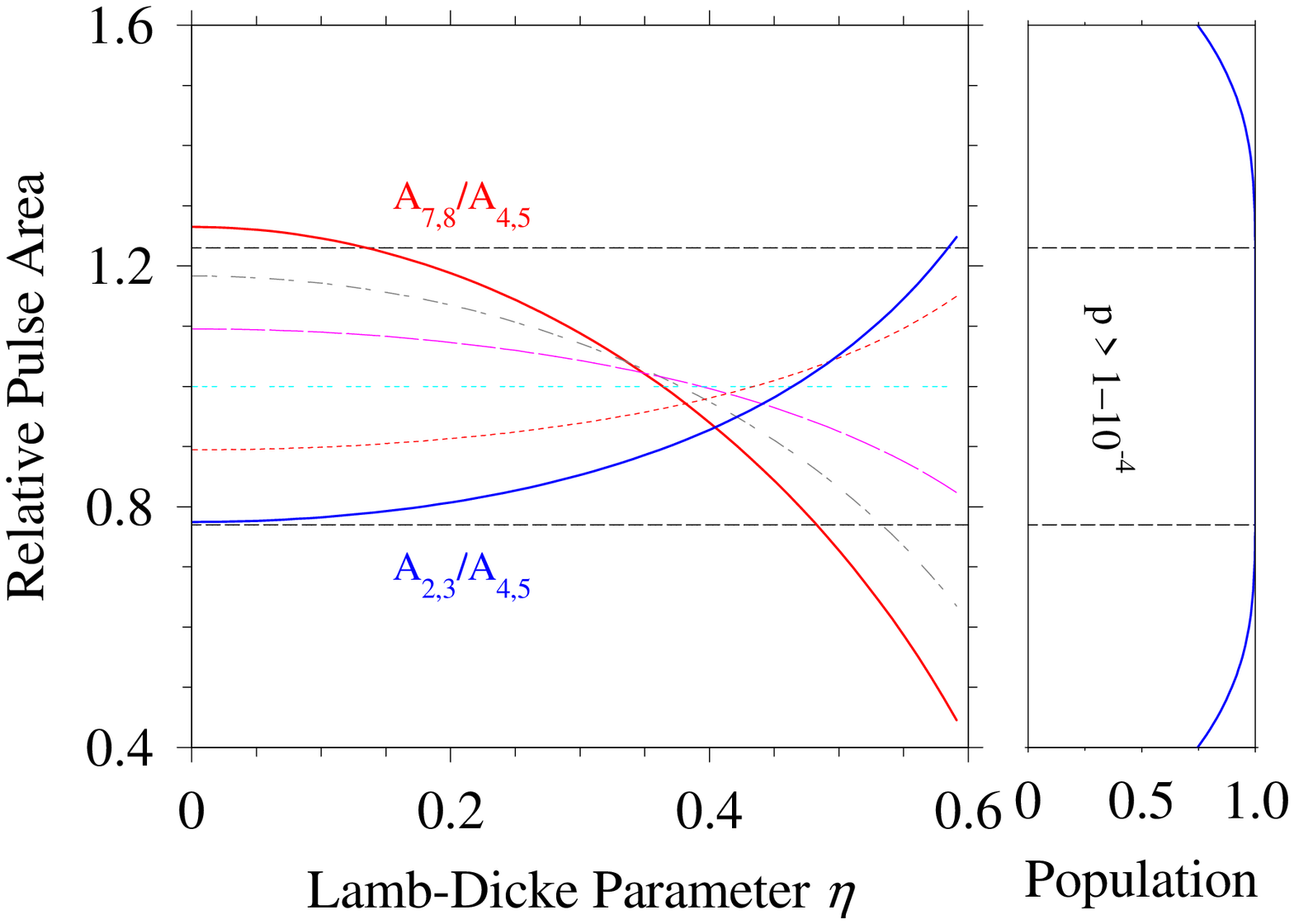}
\caption{Left: Ratios of pulse areas $A_{\phonons,\phonons+1}/A_{4,5}$ with $2\leqslant\phonons\leqslant 7$, vs the LD parameter $\eta$, needed for the Toffoli gate when the ion chain is initialized with $\phonons_0=5$ phonons. Right: Excitation profile produced by a three-component BB pulse, B$_3$, with composite phases (0, 0.65$\pi$, 0). Note that for a broad range of pulse areas around $\pi$ the transition probability $p$ for B$_3$ is very close to 1, $p \gtrsim 1-10^{-4}$ (horizontal dashed lines).
}
  \label{Fig1}
\end{figure}
%====================================================================================

\emph{Step 1.} The ion string is prepared in a common $\phonons_0$-phonon Fock state of a selected phonon mode, $\ket{0}\to\ket{\phonons_0}$, with $\phonons_0\geqslant n+1$;
 this can be achieved with $\phonons_0$ alternating blue- and red-sideband $\pi$ pulses on an ancilla ion.

%====================================================================================
\begin{figure}[t]\centering
    \includegraphics[width=0.70\columnwidth]{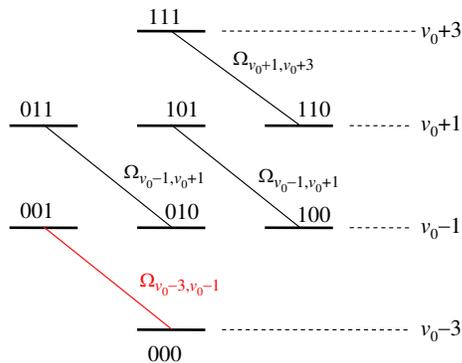}
\caption{Linkage diagram for a system of three ions, in which the third ion is addressed with a laser tuned on the second blue sideband of a selected vibrational mode.
The laser field couples states with $\phonons$ and $\phonons+2$ phonons with coupling strength dependent on $\phonons$ and the LD parameter $\eta$, Eq.~\eqref{couplings}.
}
  \label{Fig2}
\end{figure}
%====================================================================================

\emph{Step 2.} A three-component broadband (BB) composite pulse is applied on each ion on the first blue sideband, i.e. $3(n+1)$ pulses in total. This sequence is ``seen'' as a $\pi$ pulse by the transitions $\ket{0}\ket{\phonons}\fromto\ket{1}\ket{\phonons+1}$, where $\phonons_0-n-1\leqslant\phonons\leqslant \phonons_0+n$, and therefore the respective states in these transitions are inverted. This is illustrated for 3 ions in Fig. \ref{Fig1}: if the transition $\ket{0}\ket{\phonons}\fromto\ket{1}\ket{\phonons+1}$ with $\phonons=4$ phonons ``sees'' a $\pi$ pulse ($A_{4,5}\approx \pi$), then all transitions with $2\leqslant\phonons\leqslant 7$ will ``see'' (with very high fidelity) a $\pi$ pulse too.
Then the collective internal states will be dressed with a different number of phonons conditional on the number of qubits in states $\ket{0}$ and $\ket{1}$:
 a collective state with $n_k$ qubits in state $\ket{k}$ ($k=0,1$) will be dressed with $\ket{\phonons_0+n_1-n_0}$ phonons.
For 3 ions states $\ket{000}\ket{\phonons_0}$ and $\ket{111}\ket{\phonons_0}$ are mapped respectively to states $\ket{111}\ket{\phonons_0+3}$ and $\ket{000}\ket{\phonons_0-3}$,
 states $\ket{100}\ket{\phonons_0}$ and $\ket{101}\ket{\phonons_0}$ are mapped respectively to states $\ket{011}\ket{\phonons_0+1}$ and $\ket{010}\ket{\phonons_0-1}$, etc.
Thus the collective states $\ket{q_1q_2\cdots q_{n+1}}\ket{\phonons}$ group into sets with the same number of phonons,
 and the same total number of internal excitations, as shown in Fig.~\ref{Fig2}.
Note that after this step the state of each qubit is inverted, $\ket{q_k}\to\ket{1-q_k}$.

\emph{Step 3.} A $N$-component composite pulse sequence is applied on the target ion; this sequence is the core of our method.
It must act in such a way that the transition $\ket{0_10_2\cdots 0_n0_{n+1}}\fromto\ket{0_10_2\cdots 0_n1_{n+1}}$ ``sees'' an effective $\pi$ pulse and is inverted,
 while all other transitions ``see'' effectively a $0\pi$ or $2\pi$ pulse, i.e. remain either unchanged (for $0\pi$) or all acquire the same phase $\pi$ (for $2\pi$).
 (It turns out that the latter option is easier to realize.)
Such a discrimination is made possible by step 2, because the couplings $\Omega_{\phonons,\phonons+2}(t)$ depend on the LD parameter $\eta$
 and the number of phonons $\phonons$, which is different for the different sets \cite{Wineland98,Leibfried2003},
\be
\Omega_{\phonons,\phonons+2} = \frac{\Omega_0 \eta^2 e^{-\eta^2/2} L_{\phonons}^2\left(\eta^2\right)}{\sqrt{(\phonons+1)(\phonons+2)}},
\label{couplings}
\ee
where $L_{\phonons}^a(x)$ is the generalized Laguerre polynomial.
The different couplings produce different pulse areas, $A_{\phonons,\phonons+2}= \int_{t_i}^{t_f} \Omega_{\phonons,\phonons+2}(t) dt$.
Since two adjacent sets differ by 2 phonons we have to address the ions on the second blue vibrational sideband of a selected mode.

%====================================================================================
\begin{figure}[t]\centering
    \includegraphics[width=0.90\columnwidth]{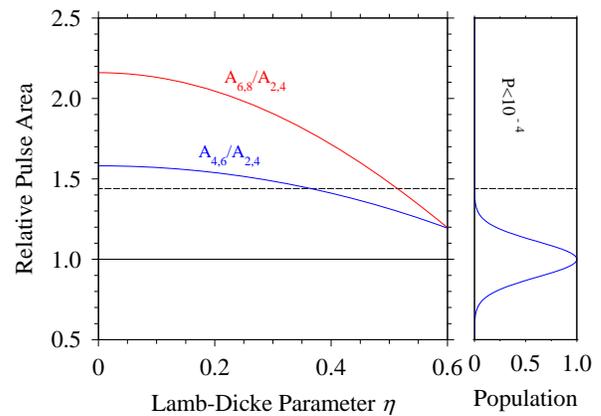}
\caption{ Left: Ratios of pulse areas $A_{6,8}/A_{2,4}$ and $A_{4,6}/A_{2,4}$ vs the LD parameter $\eta$, when the ion chain is initialized with $\phonons_0=5$ phonons.
Right: Excitation profile produced by a NB pulse N$_{13}$.
The range of gate operation is $\eta\lesssim 0.3$, where the two pulse area ratios are in the vicinity of 2.
}
  \label{Fig3}
\end{figure}
%====================================================================================

Consider a string of 3 ions needed for the Toffoli C$^2$-NOT gate.
To perform the Toffoli gate conditional on the states of ions 1 and 2 we address only ion 3 on the second blue sideband.
The laser field then produces four separate two-state systems with different couplings, as shown in Fig.~\ref{Fig2}.
We need a composite sequence, which is ``seen'' as a $\pi$ pulse by the transition $\ket{000}\ket{\phonons_0-3}\fromto\ket{001}\ket{\phonons_0-1}$
 but as a $2\pi$ pulse by the other three transitions.
Here the dependence of the couplings on the LD parameter $\eta$ is essential because the variation of $\eta$ allows us to find ranges where these conditions are fulfilled.
For an initial phonon number $\phonons_0=5$ the three subsystems in Fig.~\ref{Fig2} have couplings $\Omega_{2,4}$, $\Omega_{4,6}$ and $\Omega_{6,8}$.
The dependence of the pulse area ratios $A_{6,8}/A_{2,4}$ and $A_{4,6}/A_{2,4}$ on $\eta$ are illustrated in Fig.~\ref{Fig3}(left).
The range of values of $\eta$ where these ratios are near 2 are suitable for realization of the Toffoli gate, as indicated by the composite-pulse excitation profile in Fig.~\ref{Fig3}(right).
One can vary the initial phonon number $\phonons_0$ and produce different behavior of the relative pulse areas versus $\eta$ according to Eq.~\eqref{couplings},
 which allows one to shift the $\eta$-range for the Toffoli gate to the experimentally most convenient values.

\emph{Step 4.} We repeat step 2 \cite{comment2}; the phonon number in each collective internal state is restored to $\phonons_0$.

\emph{Step 5.} Step 1 is applied in a reverse manner; the ion chain is restored in its vibrational ground state $\ket{\phonons=0}$.

Steps 1-5 produce the transformation
\be\label{Cn-NOT-0}
\ket{\psi} \ket{0} \to [\text{C}^n\text{-NOT}\ket{\psi}] \ket{0},
\ee
with $8(n+1)$ %po-skoro 6(n+1) + 2v_0
single pulses and a composite sequence.
The number of pulses in the composite sequence from step 3 can vary, e.g., from 5 for C$^2$-NOT (Toffoli) gate to 13 for C$^6$-NOT gate, as shown below.
Because the number of pulses in the composite sequence is about $2n$,
 the total number of pulses %in the implementation of Eq.~\eqref{Cn-NOT-0}
  is about $10n$.

Our method generalizes the original idea of Monroe \etal\ \cite{Monroe1997} who constructed a CNOT gate by exploiting the $\phonons$-dependence of the LD parameter $\eta$
 to select such a ``magic'' value of $\eta$, for which the ratio of the two relevant couplings equals a rational odd/even or even/odd number.
This approach cannot be extended with a single pulse to higher gates as they require specific values of the ratios of three or more couplings.
We can satisfy these conditions in various $\eta$-ranges because composite pulses can modify the excitation profile in any desired manner.

%%%%%%%%%%%%%%%%%%%%%%%%%%%%%%%%%%%%%%%%%%%%%%%%%%%%%%%%%%%%%%%%%%%%%%%%%%%%%%%%%%%%%%%%%%%%%%%%%%%%%%%%%%%%%%%%%%%%%%%
\section{Composite sequences for C$^n$-NOT gates}\label{composite sequences}
The technique of composite pulses was introduced in nuclear magnetic resonance (NMR) \cite{NMR, Wimperis} as a powerful tool for control of spins by magnetic fields.
A composite pulse is a train of $N$ pulses with well-defined relative phases $\phase_{k}$ $(k=1,2,\dots,N)$,
 which are used as control parameters in order to compensate the imperfections of a single pulse and/or to shape up the excitation profile in a desired manner.
We have designed special composite pulses for the $C^n$-NOT gates by using a recently developed simple method \cite{Ivanov2011,Torosov}.
We use equal pulse areas $A$ and an odd number of pulses, $N=2\nc+1$, although these restrictions are not essential.
We consider symmetric ``anagram'' composite sequences,
 with phases $\phase_{k}=\phase_{N+1-k}$ $(k=1,2,\dots,\nc)$; this condition
 produces symmetric excitation profiles.
Because the overall phase of the composite sequence is irrelevant
 we set $\phase_1=\phase_{N}=0$; hence we are left with $\nc$ different phases.

%***************************************************************
\begin{table}[tb]
\begin{center}
\begin{tabular}{|l l|}
\hline
  & standard NB pulses \\ \hline
N$_5$     & (1.160; 0.580) \\
N$_9$     & (1.130; 0.820; 0.110; 1.390) \\
N$_{13}$  & (1.270; 0.440; 1.020; 0.770; 1.850; 1.730) \\
N$_{17}$  & (1.600; 0.550; 1.090; 0.890; 0.620; 1.540; 0.150; 1.570) \\
N$_{21}$  & (1.070; 0.920; 0.130; 1.830; 1.160; 0.720; 0.100; 1.520; \\
          &  0.810; 1.950) \\
N$_{25}$  & (1.750; 0.380; 1.420; 0.710; 1.070; 0.910; 0.780; 1.470; \\
          &  0.550, 1.740, 0.160, 1.650) \\
\hline
  & optimized NB pulses \\ \hline
N$_5^\opt$     & (1.190; 0.630) \\
N$_9^\opt$     & (1.157; 0.888; 0.218; 1.529) \\
N$_{13}^\opt$  & (0.585; 1.352; 0.914; 1.186; 0.020; 0.089) \\
\hline
\end{tabular}
\end{center}
\caption{Phases $(\phi_2;\phi_3;\phi_4;\ldots;\phi_{m+1})$ (in units $\pi$) for standard (N$_N$) and optimized (N$_N^\opt$) NB sequences of $N=2\nc+1$ pulses of area $A$: $A_0 A_{\phi_2}A_{\phi_3}\cdots A_{\phi_{\nc+1}}\cdots A_{\phi_3}A_{\phi_2}A_0$.
}
\label{table1}
\end{table}
%***************************************************************

Three families of composite pulse sequences are particularly important: broadband (BB), narrowband (NB) and passband (PB) \cite{Wimperis}.
The BB pulses stabilize population inversion to values $p\approx 1$ around the pulse area $\pi$ (flat-top excitation profile).
The NB pulses stabilize $p$ to values $p\approx 0$ around the pulse area $0\pi$ (or $2\pi$) (flat bottom).
The PB pulses stabilize $p$ both to values $p\approx 1$ around area $\pi$ and to values $p\approx 0$ around area $0\pi$ (or $2\pi$) (flat-top and flat-bottom).

The composite phases for the BB sequence used in steps 2 and 4 are (0, 0.65$\pi$, 0). Other examples of BB sequences can be found in \cite{Torosov}.

In step 3 all but one of the couplings must fall in the region where $p\approx 0$, hence stabilization is needed around this value;
 the remaining coupling must fall in the range where $p\approx 1$, i.e. near $\pi$.
This condition suggests to use NB pulses because of their flat-bottom excitation profiles around areas $0\pi$ and $2\pi$.
None of the existing NB composite pulses, however, produces excitation profiles with sufficiently wide bottoms to satisfy these conditions with high fidelity; for this reason we construct here new NB composite sequences.
We derive the NB phases from the conditions
\begin{equation}\label{conditions2}
[\partial^k_A U_{11}^{(N)}]_{A=2\pi} = 0 \quad (k=2,4,\dots,2\nc),
\end{equation}
with $\partial^k_A\equiv\partial^k / \partial A^k$; the skipped derivatives vanish identically for ``anagram'' sequences.
$\U^{(N)}$ is the full propagator,
$\U^{(N)} = \U_0 \U_{\phase_{2}} \U_{\phase_{3}} \cdots \U_{\phase_{m+1}} \cdots \U_{\phase_{3}} \U_{\phase_{2}} \U_0$, with
\be
\U_{\phase} = \left[ \begin{array}{cc} \cos(A/2) & i\e^{-i\phase}\sin(A/2)  \\ i\e^{i\phase}\sin(A/2)  & \cos(A/2) \end{array}\right].
\ee
Examples of NB sequences suitable for C$^n$-NOT gates are presented in Table \ref{table1}.
A typical NB excitation profile is plotted in the right frame of Fig.~\ref{Fig3}.
These NB sequences are ``all-purpose'' ones, i.e. they are suited whenever an increased selectivity of excitation, or suppression of unwanted excitation is needed,
 e.g. for local addressing in a lattice of closely spaced qubits \cite{Ivanov2011}.
We have constructed special NB sequences which optimize the performance of the C$^n$-NOT gates; they are listed in Table \ref{table1} as well.
Their phases are obtained by starting from a particular ``all-purpose'' NB pulse
 and minimizing the error in the high-fidelity region of the respective C$^n$-NOT gate in the $(\eta,A)$ control landscape.
We can always accommodate more couplings (that have to be suppressed) within the no-transition bottom of the excitation profile around $2\pi$ (and $0\pi$) by adding more pulses to the composite sequence in order to broaden this bottom.

%====================================================================================
\begin{figure}[t]\centering
   \includegraphics[width=0.9\columnwidth]{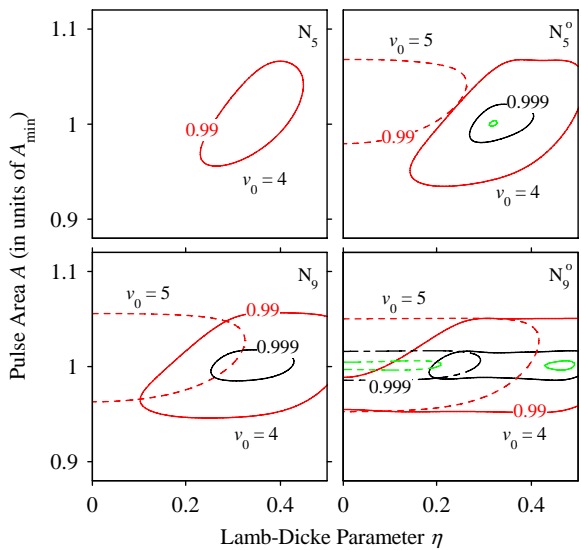}
\caption{Simulated fidelity of the Toffoli gate with NB composite sequences from Table \ref{table1} for initial phonon number $\phonons_0=4$ (solid lines) and 5 (dashed lines) vs. the LD parameter $\eta$ and the pulse area $A$.
\emph{Left:} standard NB sequences N$_5$ (top) and N$_9$ (bottom). \emph{Right:} optimized NB sequences N$_5^\opt$ (top) and N$_9^\opt$ (bottom).
The area is divided by $A_{\text{min}}=A_{\phonons_0-3,\phonons_0-1}$, that is the (smallest) pulse area which has to be ``seen'' as a $\pi$ pulse by the transition $\ket{000}\ket{\phonons_0-3}\fromto \ket{001}\ket{\phonons_0-1}$, cf. Figs.~\ref{Fig2} and \ref{Fig3}.
}
\label{fig4}
\end{figure}
%====================================================================================

After the NB sequence the state of the target ion is changed as $\ket{0}\rightarrow\e^{-i\compphase}\ket{1}$ and $\ket{1}\rightarrow\e^{i\compphase}\ket{0}$, where
\be
\compphase=(-1)^{\nc}\left(\pi/2+\phi_{\nc+1}\right)-2\sum_{k=2}^{\nc}(-1)^k\phi_k.
\ee
This phase can be compensated, if necessary, with an additional Stark pulse focused on the target ion.

%%%%%%%%%%%%%%%%%%%%%%%%%%%%%%%%%%%%%%%%%%%%%%%%%%%%%%%%%%%%%%%%%%%%%%%%%%%%%%%%%%%%%%%%%%%%%%%%%%%%%%%%%%%%%%%%%%%%%%%
\section{Simulation of C$^n$-NOT gates}\label{simulation}
We have simulated numerically different C$^n$-NOT gates $\Toff$ by using our approach presented above.
The fidelity of the simulated gate $\ToffT$ is defined as the uniform average over infinitely many random states $\ket{\psi}$:
 $F=\mathrm{mean}_\psi \left\vert \bra\psi \ToffT^\dagger\Toff \ket\psi \right\vert^2$.
Figure \ref{fig4} shows the fidelity of the Toffoli gate constructed with standard (left) and optimized (right) NB sequences from Table \ref{table1}
 for 5 (top) and 9 (bottom) ingredient pulses.
Fidelity above 99\%, and even 99.9\%, can be obtained with just 5 pulses.
Higher fidelity can be obtained with longer sequences, e.g. over 99.99\% with the N$_9^o$ pulse.
The high-fidelity $\eta$ ranges can be varied by varying the initial phonon number $\phonons_0$, as evident from the examples with $\phonons_0=4$ and 5 in Fig.~\ref{fig4}:
 a larger $\phonons_0$ shifts the high-fidelity $\eta$-range toward $\eta=0$.

%====================================================================================
\begin{figure}[tbf]\centering
    \includegraphics[width=0.9\columnwidth]{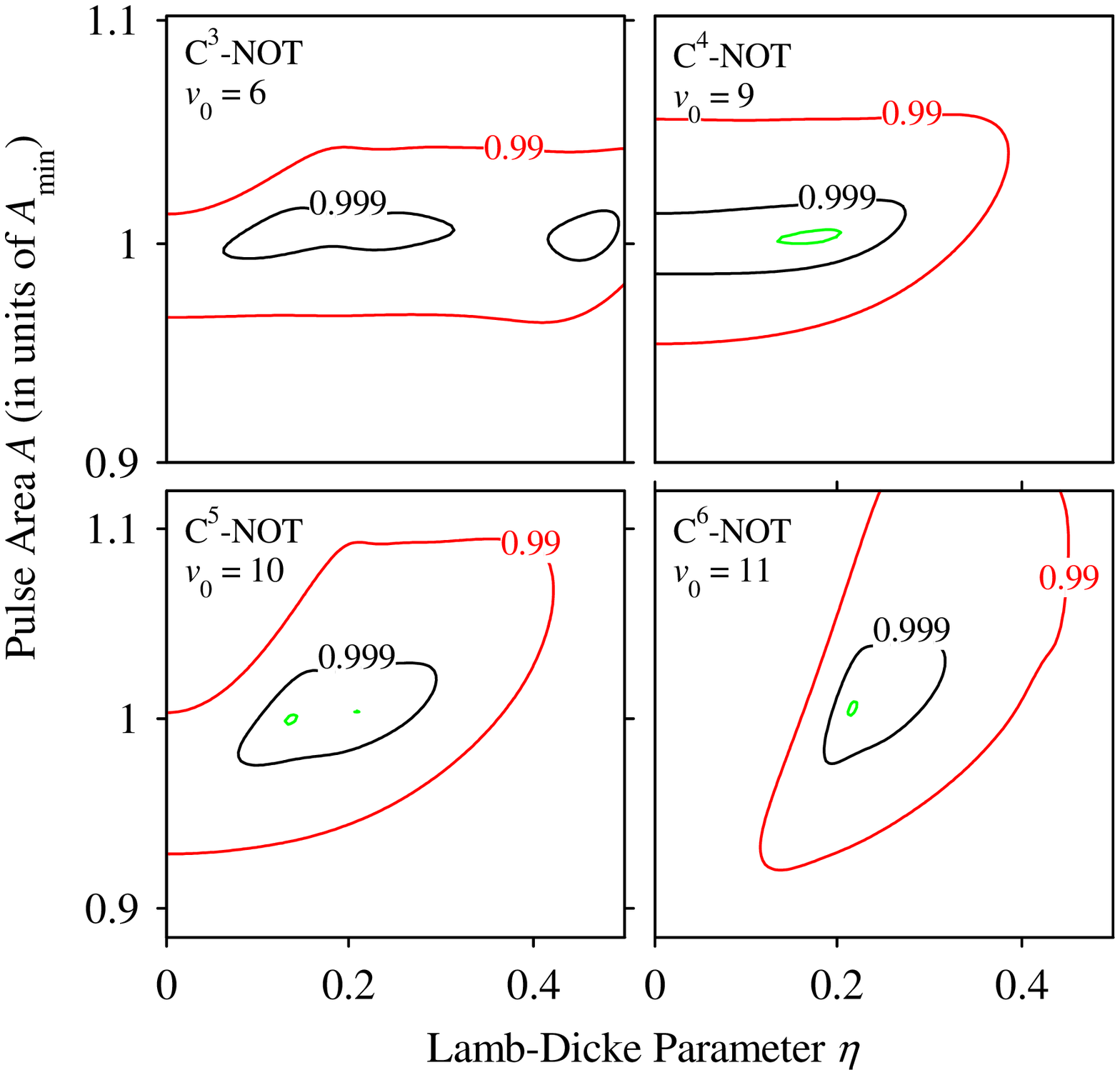}
\caption{Simulated fidelity of C$^n$-NOT gates with $n=3$ to 6 with the optimized NB sequence N$_{13}^o$ from Table \ref{table1}
 vs. the LD parameter $\eta$ and the pulse area $A$.
The area is divided by $A_{\text{min}}=A_{\phonons_0-n-1,\phonons_0-n+1}$, that is the (smallest) pulse area which has to be ``seen'' as a $\pi$ pulse by the transition
 $\ket{00\cdots 00}\ket{\phonons_0-n-1}\fromto \ket{00\cdots 01}\ket{\phonons_0-n+1}$.
Note that the same composite sequence N$_{13}^o$ implements all gates efficiently.}
  \label{fig5}
\end{figure}
%====================================================================================

We have simulated several higher-order C$^n$-NOT gates with $n=3$ to 6 using the NB sequences from Table~\ref{table1};
 the fidelities for the optimized NB pulse N$_{13}^\opt$ are shown in Fig.~\ref{fig5}.
High fidelity can be achieved in various ranges of $\eta$ by changing the initial phonon number $\phonons_0$ because the ratios of the couplings depend on both $\eta$ and $\phonons_0$;
 larger $\phonons_0$ push the high-fidelity range again toward smaller $\eta$.
In this manner, one can adjust the operation of the C$^n$-NOT gate to a range of values of $\eta$, which is most suitable experimentally.
Fidelity can always be increased, and the high-fidelity ranges can be expanded, by adding more pulses to the composite sequence, which will allow enhanced optimization.
It is remarkable that the same composite pulse N$_{13}^\opt$ implements all gates up to $n=6$ efficiently, in similar ranges of $\eta$ and $A$.
We have checked that this same pulse N$_{13}^\opt$ can be used with over 99\% fidelity for C$^n$-NOT gates up to $n=10$.
The lower gates with $n=3$ and 4 can be implemented efficiently also with the shorter pulse N$_9^\opt$ (not shown), which has been used in Fig.~\ref{fig4}.
This uniformity is strikingly different from the circuit model, in which the complexity of implementation increases rapidly with $n$,
 which makes the demonstration of most algorithms still impossible with current technology \cite{Blatt2009}.
We have verified that, for up to $n=20$, NB composite sequences of about $2n$ ingredient pulses are sufficient for efficient construction of the C$^n$-NOT gate.

%%%%%%%%%%%%%%%%%%%%%%%%%%%%%%%%%%%%%%%%%%%%%%%%%%%%%%%%%%%%%%%%%%%%%%%%%%%%%%%%%%%%%%%%%%%%%%%%%%%%%%%%%%%%%%%%%%%%%%%
\section{Discussion and conclusion}\label{conclusion}
The proposed method offers a conceptually simple and scalable implementation of C$^n$-NOT gates of trapped-ion qubits of arbitrary order $n$,
 as it requires, besides the usual initial dressing and the final undressing of the collective qubit states with phonons,
 just a single NB pulse applied on the target ion qubit.
This method uses the basic physical notion of destructive interference of unwanted transitions
 rather than quantum circuits of a vast number of concatenated one- and two-qubit gates.
The ensuing simplicity and universality make our method easily scalable to arbitrary C$^n$-NOT gates with essentially the same level of complexity.
The constructed NB sequences produce C$^n$-NOT gates with very high fidelity using relatively few ingredient pulses.
We have found that a NB composite sequence of about $2n$ pulses suffices for a C$^n$-NOT gate; including the dressing and undressing steps,
the total number of pulses is about $10n$.
%this number is similar to the number of dressing and undressing pulses in the initial and final steps.
This makes possible the creation of C$^n$-NOT gates for up to $n=15-20$ ion qubits with the existing ion trap technology \cite{Monz2011}.
We point out that the NB composite sequences are very convenient in another respect:
 they eliminate spatial imperfections of local addressing \cite{Ivanov2011}, which are often a limiting factor in experiments \cite{Blatt2009}.

We note that the use of blue-sideband pulses is not mandatory; one can use red-sideband pulses too.
One can also replace the NB composite sequences, which minimize the number of ingredient pulses,
 by PB sequences, which require a few more ingredient pulses but provide in return greater robustness against variations in the pulse area.

We point out in conclusion that the initial dressing of the qubit states with phonons and the final restoration of the phonon ground state, each of which requires about $4n$ pulses,
 can be optimized for each C$^n$-NOT gate to fewer pulses, about $n$ for each of the dressing and undressing steps \cite{IvanovUnpublished}.
The procedure described above is, however, universal (applicable to arbitrary $n$) and flexible, for it applies to large ranges of values of the LD parameter $\eta$;
 the latter feature allows one to operate outside the LD regime and hence speed up the gate operation.

%%%%%%%%%%%%%%%%%%%%%%%%%%%%%%%%%%%%%%%%%%%%%%%%%%%%%%%%%%%%%%%%%%%%%%%%%%%%%%%%%%%%%%%%%%%%%%%%%%%%%%%%%

This work is supported by the European Commission project FASTQUAST and the Bulgarian NSF grants D002-90/08 and DMU02-19/09.

%%%%%%%%%%%%%%%%%%%%%%%%%%%%%%%%%%%%%%%%%%%%%%%%%%%%%%%%%%%%%%%%%%%%%%%%%%%%%%%%%%%%%%%%%%%%%%%%%%


\begin{thebibliography}{99}
\bibitem{comment1}
This does not necessarily impose the LD regime, which is accessed when both the phonon number and the LD parameter are very small.
\bibitem{comment2}
One can shift all phases with $\pi$ relative to step 2 to compensate possible experimental errors of steps 2 and 4.
\end{thebibliography}
\end{document}